\documentclass[12pt,twoside,dvips]{article}
\usepackage{epsfig}
\usepackage{a4wide}

\newcommand{\lsim}{\rlap{\raise 2pt \hbox{$<$}}{\lower 2pt \hbox{$\sim$}}}
\newcommand{\gsim}{\rlap{\raise 2pt \hbox{$>$}}{\lower 2pt \hbox{$\sim$}}}

\newcommand{\ie}{{\it i.e.\ }}
\newcommand{\eg}{{\it e.g.\ }}

\begin{document}
\begin{flushright}
\begin{minipage}[t]{37mm}
{\bf 
hep-ph/9803496\\  
TSL/ISV-98-0193\\ 
DESY 98-035\\
March 1998} 
\end{minipage}
\end{flushright}

\vspace{2cm}
\begin{center}

{\bf \Large A model for the parton distributions in hadrons}

\vspace{20mm}

{\Large A.~Edin$^{a,}$\footnote{edin@tsl.uu.se} and 
G.~Ingelman$^{a,b,}$\footnote{ingelman@tsl.uu.se} \\ }
\vspace{3mm}
$^a$ Dept. of Radiation Sciences, Uppsala University,
Box 535, S-751 21 Uppsala, Sweden\\
$^b$ Deutsches Elektronen-Synchrotron DESY,
Notkestra\ss e 85, D-22603 Hamburg, Germany\\ 

\end{center}

\vspace{5mm}

{\bf Abstract:} 
A simple model is presented for the parton distributions in hadrons.
The parton momenta in the hadron rest frame are derived from a spherically 
symmetric, Gaussian, distribution having a width motivated by the 
Heisenberg uncertainty relation applied to the hadron size.
Valence quarks and gluons originate from the `bare' hadron, while sea
partons arise mainly from pions in hadronic fluctuations. 
Starting from a low $Q^2$ scale, the distributions are evolved 
with next-to-leading order DGLAP and give the proton structure function 
$F_2(x,Q^2)$ in good agreement with deep inelastic scattering data. 

\vspace*{1cm}


\vfill
\clearpage
\setcounter{page}{2}

The parton distributions in hadrons  play a very important role in particle
physics. The factorization theorems of QCD show that they can be used to 
calculate the cross-section for hard processes with incoming hadrons by
convoluting them with parton level cross-sections calculated  using
perturbation theory. The parton distributions are universal so that each hadron
has a unique parton structure which can be used to calculate all hard processes
involving that hadron. The parton distributions $f_i(x,Q^2)$ are interpreted as
the probability to find a parton $i$ (quark of some flavour or gluon) with a
fraction $x$ of the hadron momentum when probed by the momentum transfer $Q^2$.
The $Q^2$-dependence is very successfully described by the DGLAP
equations~\cite{DGLAP}  in perturbative QCD (PQCD). Given the input
distributions in $x$ at a scale $Q_0^2$ large enough for PQCD to be applicable,
one can calculate the distributions at any higher $Q^2$.

However, this starting $x$-shape, which depends on non-perturbative QCD
dynamics of the bound state hadron, has not yet been successfully derived from
first principles. Instead they are obtained by fitting parameterizations 
to data, in particular structure function measurements in
deep inelastic lepton-nucleon scattering (DIS), \eg the GRV~\cite{GRV},
CTEQ~\cite{CTEQ} and MRS~\cite{MRS} parameterizations.

In this Letter we present a simple theoretical model to derive the parton 
distributions from the non-perturbative dynamics confining the partons in
hadrons. The basic idea is to define the parton momentum distributions in the
hadron rest frame where they should be spherically symmetric. The shape of the
momentum distributions should be close to a Gaussian as a result of many
interactions binding the partons in the hadron. The typical width of this
distribution is a few hundred MeV from the Heisenberg uncertainty relation
applied to the hadron size. The Gaussian momentum distribution also has
phenomenological support. The Fermi motion in the proton provides the
`primordial transverse momentum', which has been extracted from deep inelastic
data and found to be well described  by a Gaussian distribution of a few
hundred MeV width~\cite{kt}. However, this width depends on at what $Q^2$ scale
it is extracted, since perturbative QCD effects from emission of partons in the
initial state may also contribute.

This approach is not intended to provide the full wave function for the hadron,
but only the four-momentum $k$ of a single probed parton. All other partons are
treated collectively as a single remnant with  four-momentum $r$, which
corresponds to integrating out all other information in the hadron wave
function. 

\begin{figure}[htb] 
\begin{center}
\epsfig{file=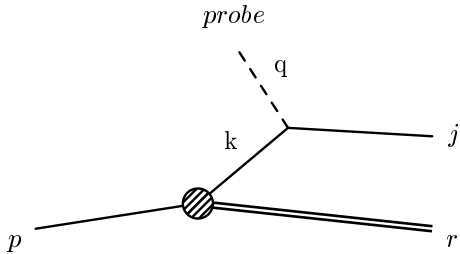} 
\end{center}
\caption{The probe $q$, which can be either a photon or a parton initiating a 
DGLAP evolution chain, probes a parton $k$, giving a scattered parton $j$ and 
a remnant system $r$.} 
\label{fig:dis} 
\end{figure}

The arguments above define only the three-momentum of the probed parton.  The
energy component does not have the same simple connection to the Heisenberg 
uncertainty relation. For simplicity, we assume it to be the current mass  of
the parton plus a Gaussian variation with the same width as for the 
three-momentum components. Thus, partons can be off-shell at this soft scale 
as expected from the soft binding interactions. This means a parton fluctuation
life-time corresponding to the hadron radius. The parton is probed at the scale
$Q_0^2$ supplied either by a virtual photon directly or indirectly through the
starting point of a DGLAP evolution chain. The scale $Q_0^2$ must be
sufficiently large, such that the parton can be considered `free' in the
interaction. Fig.~\ref{fig:dis} illustrates the basic process and defines the
relevant  four-momenta. 

The coordinate system is chosen with the negative $z$-direction along the
probe. The momentum fraction $x$ of the parton is then defined as the {\it
light-cone fraction} along the positive direction, \ie $k_+=x p_+$, which is
equivalent to $k_z=x p_z$ in a frame where $p_z$ is large. The light-cone
fraction $x$ is invariant with respect to boosts along the $z$-axis. It is
only  possible to scatter on partons that give an allowed final state. The
scattered parton must have a  mass-squared in the range  $0<j^2<W^2$, where $W$
is the invariant mass of  the hadronic system. Furthermore, the hadron remnant
must have a sufficient mass  to contain the remaining partons, \ie $r^2>\sum_i
m_i^2$ where the sum is over all partons in the remnant whose internal dynamics
is  neglected. At large energy $j^2=k_+q_-$, so that $j^2>0$ is equivalent to
$x>0$, and $j^2<W^2$ means that $x<1$. For cases where there is only light
quarks in the remnant it is enough to require $r^2>0$.

The parton density distributions are calculated numerically from the model 
using a Monte Carlo technique. The momentum components of the parton  (to be
probed) is chosen from a Gaussian distribution, as described above,  which
provides the vector $k$. The four-momentum $p$ of the hadron at rest is simply
given by its mass.  The four-momentum $q$ of the probe is given by its
virtuality $Q_0^2$ and $q_-$. The former is a free parameter determined from
data (see below), but expected to be of order 1 GeV$^2$, whereas $q_-$ must be
large  (but the exact value is not important) to ensure that the mass of the 
produced hadronic system is above the resonance region.  The internal dynamics
of the remnant (which is not measured) can then be  neglected  and the
probability for hadronization is unity. The four-momenta $j$ and $r$ are then
calculated from energy-momentum conservation and the exact  kinematical
constraints checked.  If they are fulfilled, the light-cone fraction $x$ of the
parton is added to the parton distribution. Iterating this procedure gives the
parton density distributions $f_i(x)$ at $Q_0^2$.

As an example of how the parton distributions are obtained in the model we
take the proton and first consider only the valence quarks. It is obvious
that the distributions $u_v(x)$ and $d_v(x)$ must satisfy the normalization
conditions $\int_0^1 u_v(x)dx=2$ and $\int_0^1d_v(x)dx=1$, to get the correct
quantum numbers for the proton. In addition, there must be a gluon distribution
$g(x)$ to represent the colour field and account for the fraction of the proton
momentum carried by electrically neutral partons. 
Since the gluons are confined in essentially the
same region (\ie the proton) as  the quarks, they are assumed to have the same
basic Gaussian shape as the valence quarks. The normalization of the gluon
density is given by the momentum sum rule  $\int_0^1 \left( xu_v(x) + xd_v(x) +
xg(x)\right) dx=1$.
 
The parameters of the model for the valence partons are the widths ($\sigma$)
of the Gaussian distributions for $u_v$, $d_v$ and $g$, whereas their
normalization is given by the number and momentum sum rules. The widths are, as
discussed, expected to be a few hundred MeV, but since they cannot be predicted
accurately we treat them as free parameters and obtain their values by fitting
to structure function data as described below. The resulting Gaussian widths
are $\sigma_u=180, \: \sigma_d=150,\: \sigma_g=135\:$ MeV which are reasonable
considering that the proton radius is $\sim 200$ MeV$^{-1}$. Since $\sigma$
applies in each dimension, one obtains a  two-dimensional primordial transverse
momentum with  $\langle k_\perp^2 \rangle = 2 \sigma^2$ in basic agreement with
data  \cite{kt}.

\begin{figure}[tb]
\begin{center}
\vbox{
\epsfig{file=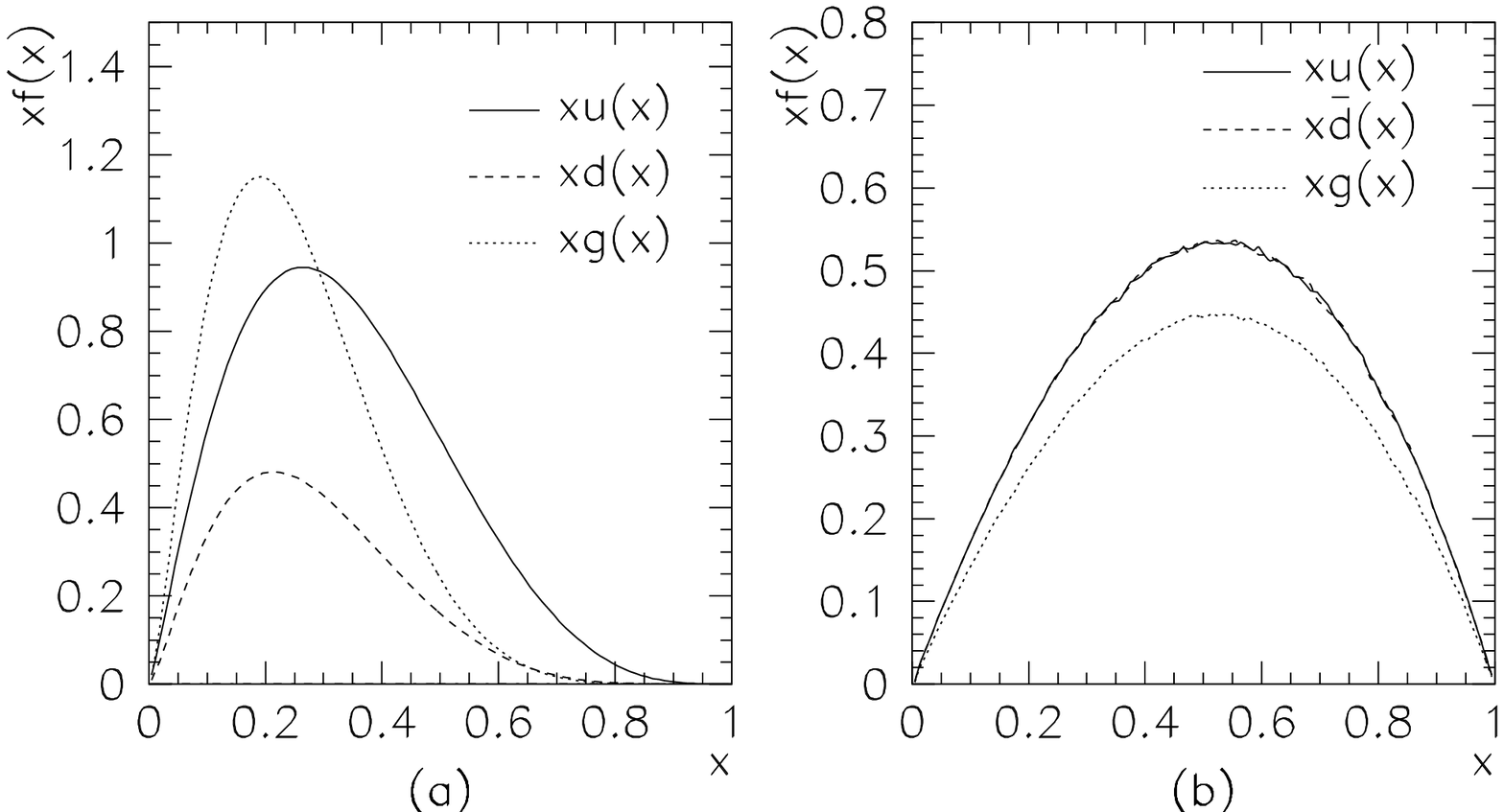,width=12cm}
\epsfig{file=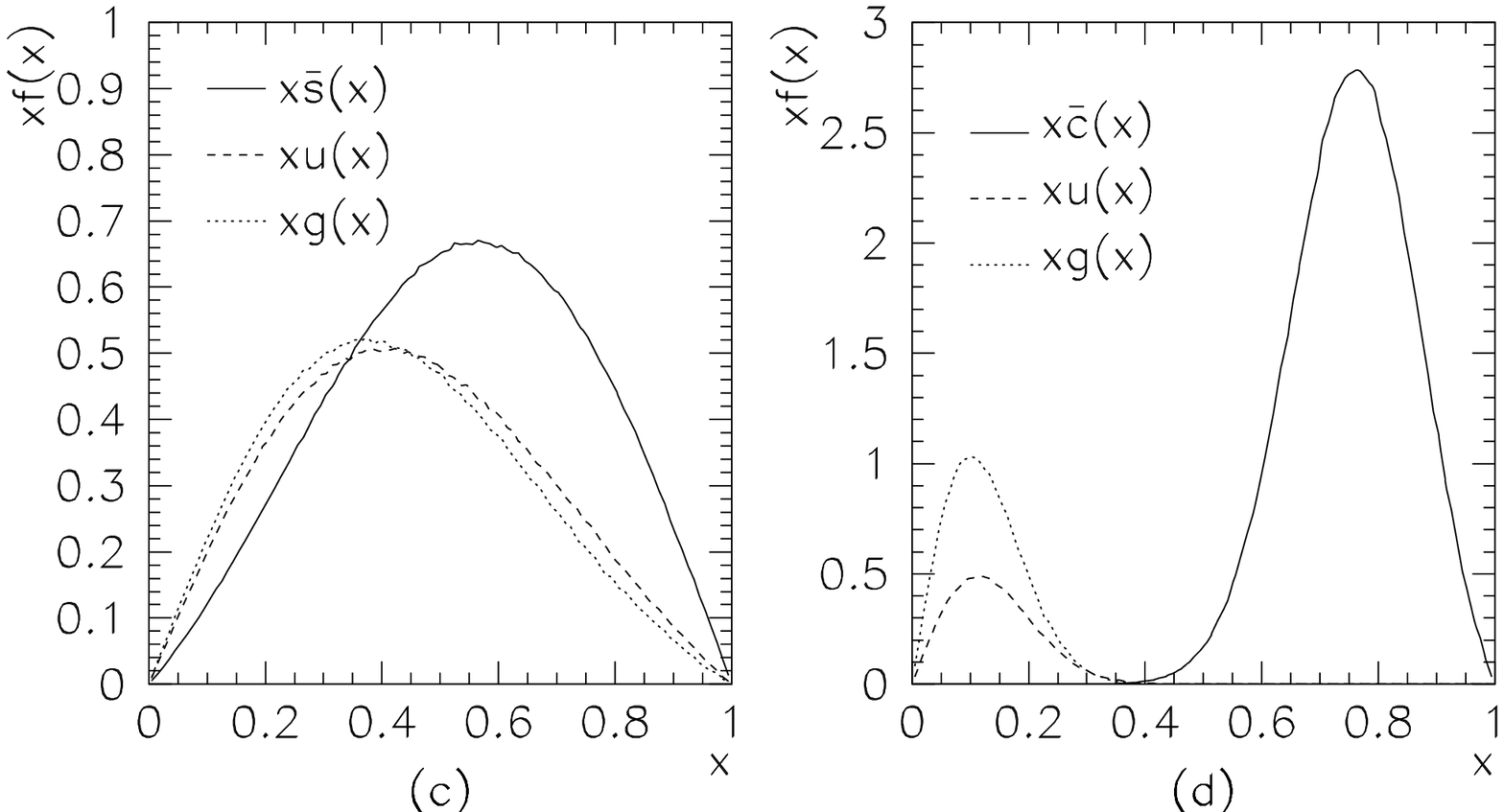,width=12cm}
}
\end{center}
\caption{The valence quark and gluon distributions obtained from the model
applied to {\bf (a)} the proton, {\bf (b)} the pion, 
{\bf (c)} the strange meson $K^+$, 
{\bf (d)} the charm meson $\bar{D}^0$. 
The Gaussian widths used are 135 MeV for gluons and 150 MeV for 
$q$ and $\bar{q}$, except $\sigma_u = 180$ MeV in (a).
}
\label{fig:valence}
\end{figure}

The momentum-weighted distributions $xf_i(x)$ for the proton as obtained
from the model are displayed in 
Fig.~\ref{fig:valence}a. The distributions look like conventional
valence quark parameterizations at a low  $Q^2$ scale and are similar to the
GRV parameterization~\cite{GRV} which is also defined at a low scale.
The proton momentum is in our case carried to $43\%$, $18\%$, $39\%$ by 
$u_v$, $d_v$ and gluons, respectively, and the integrated gluon number 
density is $\int_0^1 g(x)dx=2.4$. 

Following the same line of reasoning as for the proton, the model provides 
the parton distributions for other hadrons. Although different hadrons may 
have somewhat different sizes, we assume as an approximation the same widths 
of the Gaussian distributions for quarks and gluons as in the proton.  
However, if there are more than one quark of the same kind, \eg $u$--quarks in
the proton, it is a separate parameter which may reflect the slightly reduced
available region due to the Pauli principle. The $u$-quark distribution in the
proton is
indeed found to have a $\sim 20\%$ larger width corresponding to a slightly
smaller effective size. 

The resulting valence quark and gluon distributions in the pion are shown in
Fig.~\ref{fig:valence}b and found to be similar to normal parameterizations of
parton densities in the pion such as GRV \cite{GRVpi}. Applying the model to
hadrons containing heavier quarks we obtain the distributions shown in
Figs.~\ref{fig:valence}cd using $m_s=0.2$ GeV and $m_c=1.4$ GeV.  In
particular, the hard charm quark distribution is here a result of the charm
quark mass in the applied kinematical constraints discussed above.  

The reason why the quark distributions peak around $1/3$ in the proton and
$1/2$ in the pion (see Fig.~\ref{fig:valence}ab) has in this  model nothing to
do with having three or two valence quarks, which is anyhow dubious since one
is then neglecting the substantial fraction of the hadron momentum carried by
gluons. Instead the peak of the valence distributions depend on the ratio of
the Gaussian width (or inverse hadron size) and the hadron mass, but it is also
influenced by the kinematical contraints.  A large quark mass has a substantial
effect as illustrated in Figs.~\ref{fig:valence}cd.

To illustrate the shapes of the valence quark and gluon distributions they are
fitted to the functional form $f(x)=N x^a (1-x)^b$ that are often used to 
describe valence distributions. The results for the proton and the pion 
($\pi^+$) are
\begin{eqnarray}
\label{eq:valence} 
p:& & \!\!\!\!\! xu(x) = 13 x^{1.2} (1-x)^{3.4} \: , \: 
xd(x) = 13 x^{1.4} (1-x)^{5.0}  \: , \:  
xg(x) = 47 x^{1.4} (1-x)^{6.2}  \\
\pi^+:& & \!\!\!\!\! xu(x) = 2.3 x^{1.1} (1-x)^{1.0} \: , \: 
x\bar{d}(x) =  2.3 x^{1.1} (1-x)^{1.0} \: , \:  
xg(x) = 1.9 x^{1.1} (1-x)^{1.0} 
\end{eqnarray} 
It is interesting to note that the powers of the $(1-x)$ factors 
are quite similar to the ones in parton parametrizations. Their values are 
often motivated by counting rules~\cite{counting_rules}, although this does 
not generally work quite well since the $u_v$ and the $d_v$ in the 
proton have different powers and $g$ in the pion has the same power as the
valence quarks.

\begin{figure}[tb]
\begin{center}
\epsfig{file=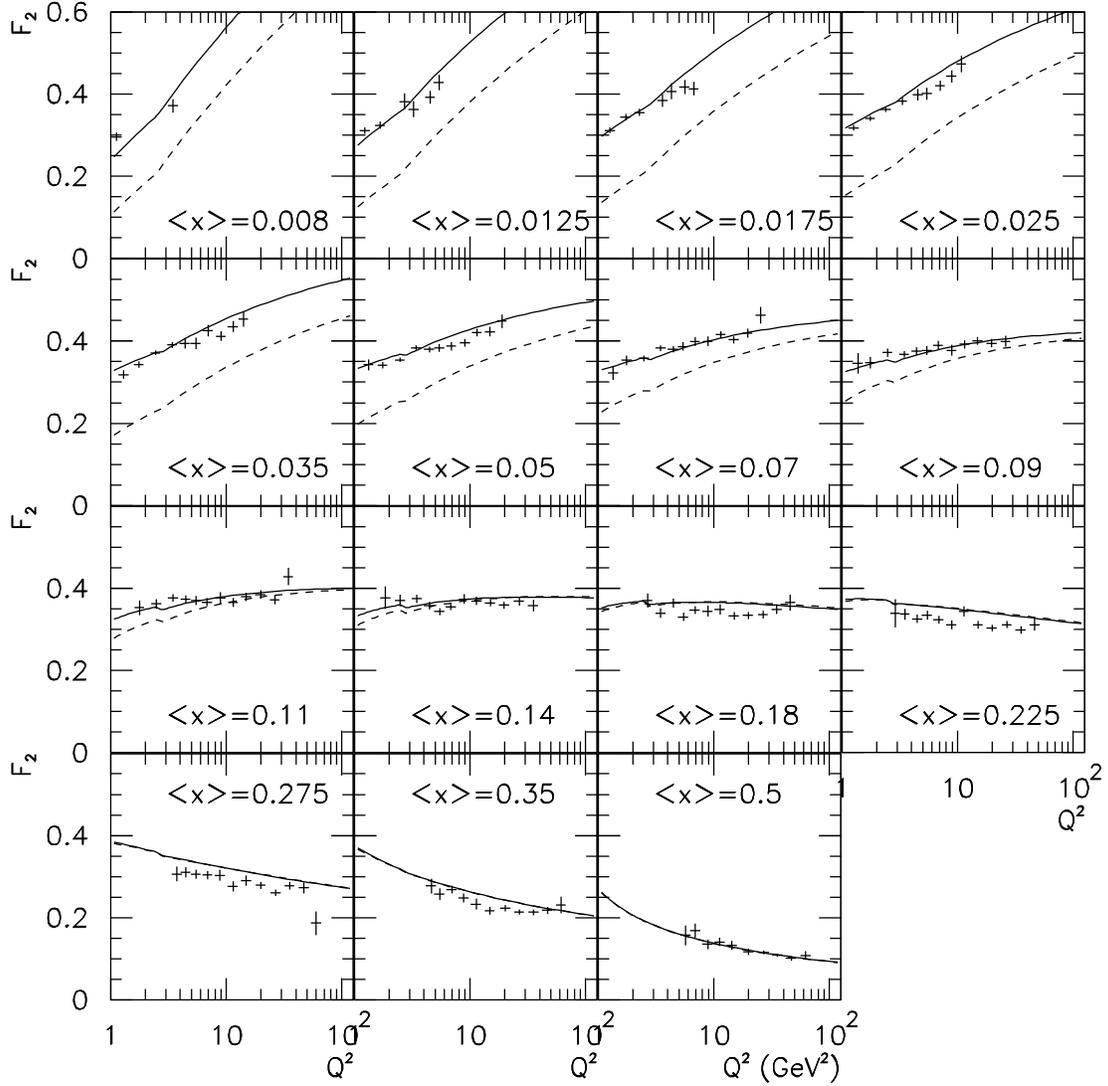,height=14.5cm}
\end{center}
\caption{The DIS structure function $F_2$ versus $Q^2$ in bins of $x$. 
Fixed target NMC data~\cite{NMC-data} compared to the model starting 
from only valence quarks and gluons (dashed) and including also a sea 
quark component (full). (The small break in the curves at $Q^2\sim m_c^2$ 
is due to the charm threshold.)}
\label{fig:nmc}
\end{figure}

Given the valence distributions of the proton, we apply QCD evolution starting
from a low staring scale $Q_0=0.6-1.0$ GeV and evolve to higher $Q^2$ to make a 
comparison with data possible. The evolution is performed using the CTEQ
program~\cite{CTEQ} which solves the next-to-leading-order (NLO) DGLAP
equations in the $\overline{MS}$ scheme. The varying effective number of quark
flavours ($n_f$) with $Q^2$ is here taken into account using the standard
procedure with $\Lambda^{(n_f)}$. For convenience, the starting distributions 
in Fig.~\ref{fig:valence}a are fitted to the shape 
$xf(x,Q_0^2)=N x^a (1-x)^b (1+cx^d)$, 
that is also used by the CTEQ collaboration in their parton distribution
fits. The evolution then gives the distributions $xf_i(x,Q^2)$ in NLO
$\overline{MS}$ scheme.

\begin{figure}[tb]
\begin{center}
\epsfig{file=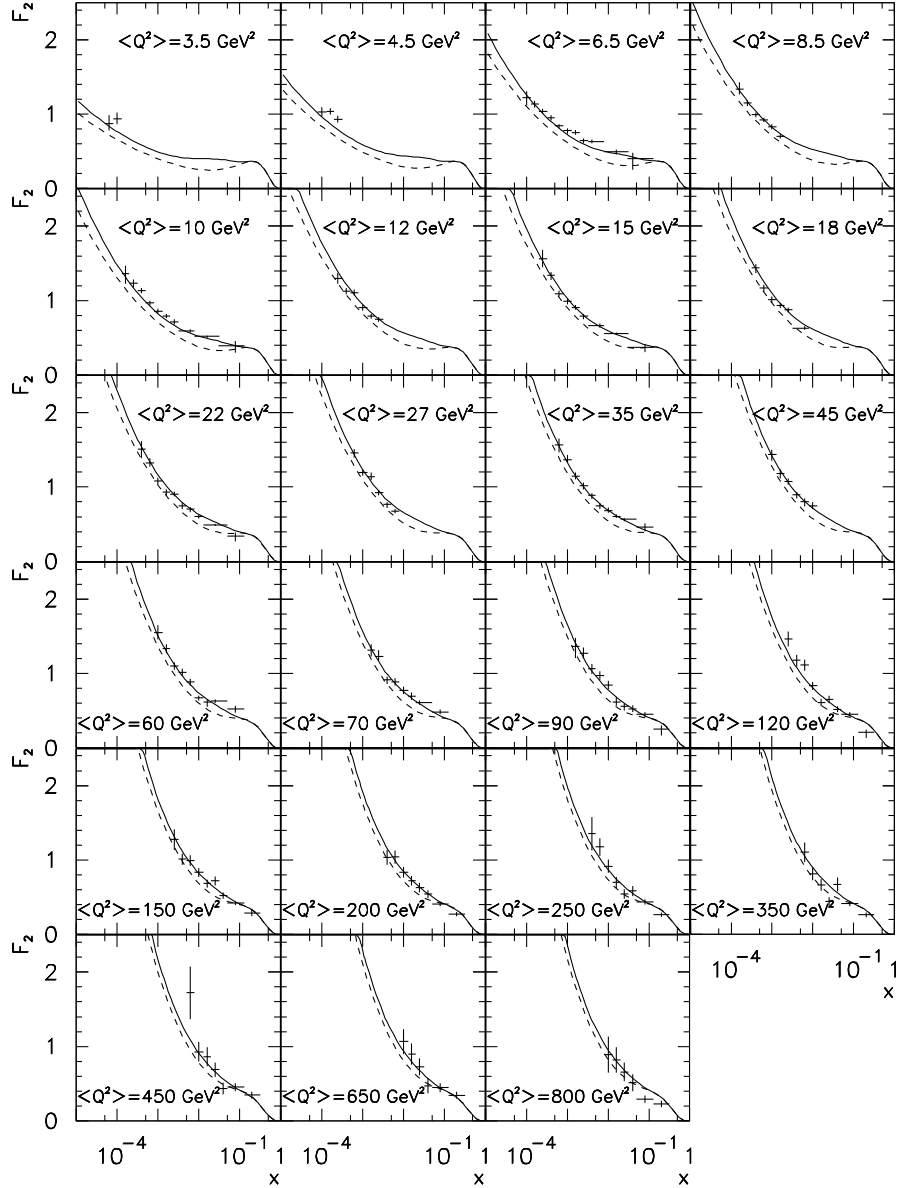,height=16cm}
\end{center}
\caption{The DIS structure function $F_2$ versus $x$ in bins of $Q^2$. 
HERA $ep$ collider data from ZEUS~\cite{Zeus-data} compared to the model 
starting from only valence quarks and gluons (dashed) and including also 
a sea quark component (full).}
\label{fig:zeus}
\end{figure}

From these we calculate the structure function $F_2(x,Q^2)$ in NLO and compare
with experimental data from fixed target muon scattering (NMC \cite{NMC-data})
and the HERA $ep$ collider (ZEUS \cite{Zeus-data}) as shown in
Figs.~\ref{fig:nmc} and ~\ref{fig:zeus}. A fit to this data is made by varying
the model parameters resulting in the numerical values for the Gaussian widths
mentioned above. When varying $\Lambda^{(5)}$ between $150$ and $300$ MeV, we
find that these widths are essentially constant (within errors), but the
cut-off parameter $Q_0$ varies linearly with $\Lambda$ as expected since only
their ratio enters in the QCD evolution. We take $\Lambda^{(5)}=0.23$ GeV,
corresponding to the measured value of $\alpha_s(M_Z)$, and get $Q_0=0.85$ GeV, which is a quite reasonable value for the PQCD cut-off. 
The $F_2$ data for the proton is most sensitive to $Q_0^2$, the $u_v$ width 
and the sea distributions, whereas the $d_v$ and the gluon widths are less 
constrained. They could be determined better by including data from, 
\eg, neutrino DIS, Drell-Yan and prompt photon production.

Our simple model with only valence quarks and gluons gives a surprisingly good
overall result, shown by the dashed lines in Figs.~\ref{fig:nmc}
and~\ref{fig:zeus}. However, the resulting structure function is too low at
small $x$ and small $Q^2$; in particular compared to the NMC data in the
region $0.008\lsim x \lsim 0.1$. Small $x$ at large $Q^2$ is not that bad
compared to HERA data, since it is strongly influenced by the QCD evolution.
This deficiency at small $x$ is an indication for the need of a sea quark
contribution already at the starting scale $Q_0^2$. This has also been noted by
GRV~\cite{GRV} and they introduced sea quark parameterizations at their low $Q^2_0$ scale to be able to fit data.

We do not want to introduce some {\it ad hoc} parameterization of sea quark
distributions to solve this problem, but rather extend our model in a natural
way to give a prescription for the sea distributions. Since our model is based
on quantum fluctuations in the proton, we consider what  fluctuations that are
most important in the non-perturbative region at  scales below $Q_0^2$. 
It seems appropriate that one should use a quantum  mechanical basis
of hadronic states and consider hadronic fluctuations.  Since the pion mass is
so small, fluctuation with virtual pions should dominate. These will have 
life-times of the order $\sim 1/m_{\pi}$, which is similar  to the widths of
the parton momentum distributions and much longer than the  time-scale $1/Q_0$
of the probe. 

Therefore, one should  consider the proton wave function as an expansion in the
hadronic Fock states containing a pion, \ie 
\begin{equation} 
\label{eq:Fockstate}  
|p\rangle  = a_0 |p\rangle  + a_1 |p\pi^0\rangle  +  a_2 |n\pi^+\rangle + 
             a_3 |\Delta\pi\rangle + \ldots,
\end{equation}
The scattering on partons in such a pion will then generate sea quark and 
gluon distributions. Scattering on the baryons in these fluctuations can be 
neglected as a first approximation,  
since they give only smaller corrections to the valence distributions in the 
dominating `bare' proton, \ie $a_0 |p\rangle $. 
The fluctuation into $|\Delta\pi\rangle$
states are less probable due to the  higher $\Delta$ mass and can at first be 
neglected. We note that already the fluctuations into 
$|p\pi^0\rangle$ and $|n\pi^+\rangle$ leads to a breaking of the $u$-$d$ 
flavour symmetry in the sea, which might explain the observed difference of 
the $\bar{u}$ and $\bar{d}$ sea quark density parameterizations \cite{udsea}.
However, this is not taken into account in this first study, where a symmetric
sea is obtained using for simplicity pion isospin symmetry or effectively 
considering only the $|p\pi^0\rangle$ term. 

The parton distributions of the pion follow from the model as described above.
In addition one must specify the momentum distribution of the pion fluctuations
in the proton. This is derived using the same arguments as for the partons, 
\ie using a spherically symmetric, Gaussian momentum distribution in
the proton rest frame. However, the width is now expected to be of order tens
of MeV based on the typical momenta of pions in the virtual pion cloud around a
proton or in a nucleus.  
Again, we treat this width as a free parameter which is fitted to data.

The normalization is in principle given by the probability amplitude
coefficients $a_i$ in eq.~(\ref{eq:Fockstate}). 
These are partly given by Clebsch-Gordan coefficients, but also depend on   
non-perturbative dynamics that cannot be calculated from first principles in 
QCD. Hence, we represent them by a free parameter for the total amount of the 
proton momentum that is carried by the sea partons generated from the pions. 

\begin{figure}[tb]
\begin{center}
\epsfig{file=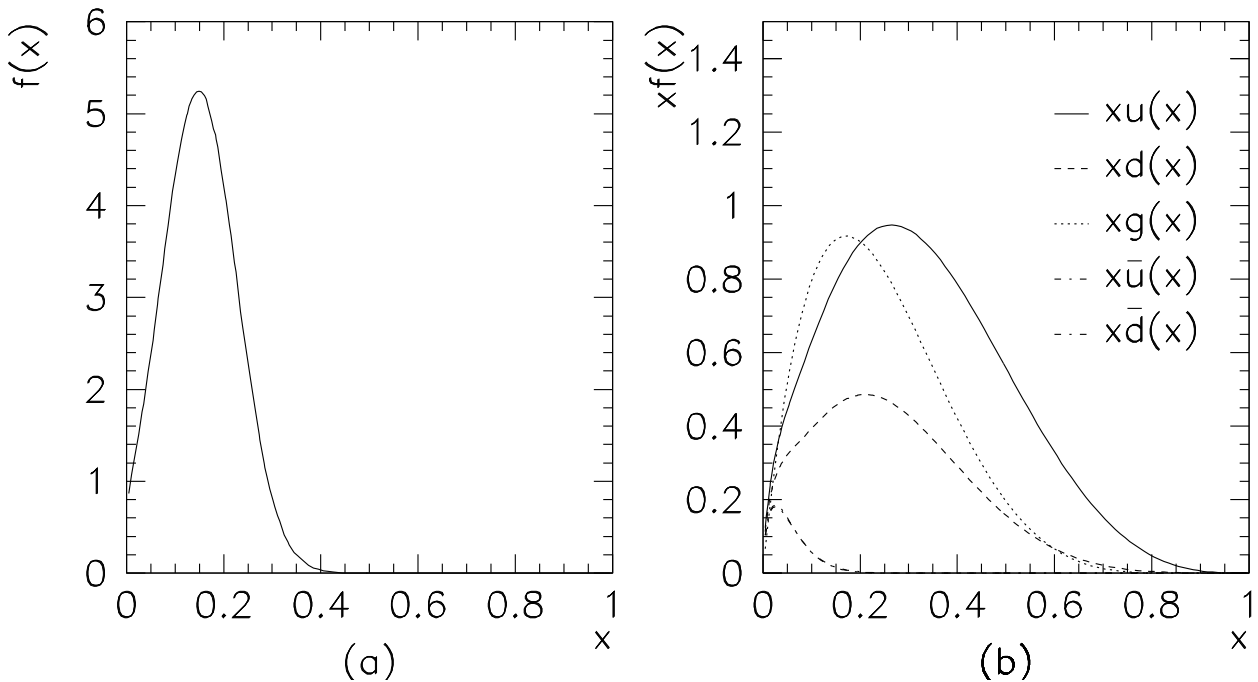,height=10cm}
\end{center}
\caption{{\bf (a)} The model result for the momentum fraction $x_{\pi}$ of the proton carried by a pion, and {\bf (b)} the valence and sea parton distributions of the proton. The quark and gluon sea are generated from pion fluctuations 
(with the sea included in the $u$, $d$ and $g$ distributions).}
\label{fig:pionsea}
\end{figure}

Applying the kinematical constraints as above, one obtains the pion momentum
distribution as shown in Fig.~\ref{fig:pionsea}a,  where $x_\pi$ is the
light-cone fraction of the proton momentum carried by the pion. One should 
note that this distribution is softer than the pion flux factor in
Regge phenomenology which has a broad maximum at $x_{\pi}= 0.2-0.5$ 
\cite{Reggepion}.
There is no reason they should be the same, since the first case relates to a
low energy quantum fluctuation whereas the Regge approach is for high energy
hadron interactions, in particular diffractive-like interactions with a
high energy leading particle produced. 

The proton sea quark distributions are then obtained by folding the pion 
momentum distribution in the proton with the valence quark and gluon
distributions in the pion. Applying the same fitting procedure of the model to
the data in Figs.~\ref{fig:nmc} and~\ref{fig:zeus}, we obtain the pion
parameters $\sigma_{\pi}=52$ MeV and that $7.7\%$ of the proton momentum is
carried by the sea partons. This width is of the expected  magnitude and the
amount of sea is similar to the GRV parameterization~\cite{GRV92}. The
resulting parton distributions, including the sea, in the proton at $Q_0=0.85$ GeV are shown in Fig.~\ref{fig:pionsea}b.

The evolved parton distributions, including the sea from pion fluctuations,
provide a quite good description of the structure function data in 
Figs.~\ref{fig:nmc} and~\ref{fig:zeus}. The parameter values of the model,
which are collected in Table~\ref{tab:parameters}, are correlated and the
minimum not very well-defined, especially for $\sigma_d$ and $\sigma_g$.
Therefore, some variations of the parameter values can result in essentially
equally good fits. The resulting $\chi^2$ is about 2 per degree of freedom,
which is not as good as in standard parton density parameterizations, such as
GRV \cite{GRV}, CTEQ \cite{CTEQ} and MRS \cite{MRS}. However, these have many
more parameters and do not provide any physical model for the non-perturbative
input parameterizations.

\begin{table}[h]
\begin{center}
\begin{tabular}{|l|l|l|l|l|l|}
\hline 
 $Q_0$ & $\sigma_u$ & $\sigma_d$ & $\sigma_g$ & $\sigma_{\pi}$ & $N_{sea}$ \\
\hline 
 850 MeV & 180 MeV & 150 MeV & 135 MeV & 52 MeV & 7.7 \%  \\
\hline
\end{tabular}
\end{center}
\caption{The model's parameter values obtained from the fit using 
$\Lambda_{\overline{MS}}^{(5)}=230$~MeV.}
\label{tab:parameters}
\end{table}

The virtue of our parton distributions is that they are derived from a simple
physical model with few assumptions and few free parameters. Although we have
determined the parameter values from data, it is very comforting to note that
their values agree well with the definite expectations based on the model. 

The model works quite well for describing the proton $F_2$ in DIS down to 
$Q^2 \simeq 1$ GeV$^2$, as seen in Fig.~\ref{fig:nmc}. At still smaller $Q^2$,
and smaller $x$ at HERA energies, the DIS formalism and the partonic 
interpretation of $F_2$ are not straightforwardly applicable. 
Photoproduction and resolved photon processes must here be included in
the theoretical description of the data, in particular for small $Q^2$ 
and $x$ at HERA. More generally, parton distributions can only be used 
when the hadron is resolved by a sufficiently large scale. 
Hence, for $Q^2<Q^2_0$ our parton distribution model can only be applied 
for processes with some other hard scale, \eg jets in photoproduction, 
and not to obtain the total photon-proton cross-section or $F_2$ extracted 
from it. 

In this Letter we have presented a model to describe the parton distributions
in hadrons, in an attempt to understand the soft interactions
which confine the quarks and gluons to hadrons. Normally the parton
distributions are obtained by fitting a parametrization to experimental data,
giving little insight into the physical processes involved.

Our model has two main contributions to the parton distributions. Valence
quarks and gluons come directly from the `bare' hadron, \eg $|p\rangle$ for the
proton, while sea quarks and `sea' gluons come from hadronic fluctuations,
mainly with pions such as $|p\pi^0\rangle$. This provides a one-to-one 
correspondence between the Fock state expansion of the hadronic wave-function
of the proton and the partonic structure in terms of valence and sea partons.
Although the model is quite simple in this first attempt, it provides parton
densities which, when evolved with standard PQCD, successfully fits structure
function data over a wide range in $x$ and $Q^2$. 

The model straight-forwardly gives the parton densities in other hadrons. Of
course, this must be tested by comparing to measurements, \eg the parton
distributions for the pion. Considering also proton fluctuations into charm, 
\eg $|\Lambda_c^+ \overline{D}^0\rangle$, would lead to a charm quark component
in the proton derived in a different way compared to the intrinsic charm model
\cite{intrinsic}. Thus, our simple model not only describes the parton
distributions in the proton well, it also has interesting features that deserve
further investigations. 

{\bf Acknowledgements.}
We are grateful to A.\ Vogt for providing the program calculating $F_2$ 
in NLO and J.\ Rathsman for many useful comments. We thank B.\ Andersson
and T.\ Sj\"ostrand for a critical reading of the manuscript.

\end{document}